\documentclass[%
aip,
amsmath,amssymb,
preprint,%
]{revtex4-1}

\usepackage{graphicx}
\usepackage{dcolumn}
\usepackage{bm}

\usepackage[utf8]{inputenc}
\usepackage[T1]{fontenc}
\usepackage{mathptmx}
\usepackage{etoolbox}
\usepackage{xcolor}
\newcommand{\CORR}[2]{#1} 

\draft 

\begin{document}


\title{Statistics of drops generated from ensembles of randomly corrugated ligaments}


\author{Sagar Pal}
\affiliation{Sorbonne Universit\'e, Institut Jean Le Rond d’Alembert, UMR 7190, Paris, France}
\affiliation{Centre National de la Recherche Scientifique, UMR 7190, Paris, France}

\author{César Pairetti}
\email[]{paire.cesar@gmail.com}
\affiliation{Sorbonne Universit\'e, Institut Jean Le Rond d’Alembert, UMR 7190, Paris, France}
\affiliation{Universidad Nacional de Rosario, Rosario, Argentina}

\author{Marco Crialesi-Esposito}
\affiliation{Department of Engineering ‘Enzo Ferrari’, University of Modena and Reggio Emilia, 41125 Modena, Italy}
\affiliation{Sorbonne Universit\'e, Institut Jean Le Rond d’Alembert, UMR 7190, Paris, France}

\author{Daniel Fuster}
\affiliation{Sorbonne Universit\'e, Institut Jean Le Rond d’Alembert, UMR 7190, Paris, France}
\affiliation{Centre National de la Recherche Scientifique, UMR 7190, Paris, France}

\author{St\'ephane Zaleski}
\affiliation{Sorbonne Universit\'e, Institut Jean Le Rond d’Alembert, UMR 7190, Paris, France}
\affiliation{Centre National de la Recherche Scientifique, UMR 7190, Paris, France}
\affiliation{Institut Universitaire de France, Paris, France}


\date{\today}

\begin{abstract}
	The size of drops generated by the capillary-driven disintegration of liquid ligaments plays a fundamental role in several important natural phenomena, ranging from heat and mass transfer at the ocean-atmosphere interface to pathogen transmission. The inherent non-linearity of the equations governing the ligament destabilization leads to significant differences in the resulting drop sizes, owing to small fluctuations in the myriad initial conditions. 
	Previous experiments and simulations reveal a variety of drop size distributions, corresponding to competing underlying physical interpretations. Here, we perform numerical simulations of individual ligaments, the deterministic breakup of which is triggered by random initial surface corrugations. The simulations are grouped in a large ensemble, each corresponding to a random initial configuration. The resulting probability distributions reveal three stable drop sizes, generated via a sequence of two distinct stages of breakup. Four different distributions are tested, volume-based Poisson, Gaussian, Gamma and Log-Normal. Depending on the time, range of droplet sizes and criteria for success, each distribution has successes and failures. However the Log-Normal distribution roughly describes the data when fitting both the primary peak and the tail of the distribution while the number of droplets generated is the highest, while the Gamma and Log-Normal distributions perform equally well when fitting the tail. The study demonstrates a precisely controllable and reproducible framework, which can be employed to investigate the mechanisms responsible for the polydispersity of drop sizes found in complex fluid fragmentation scenarios. 
\end{abstract}


\maketitle

\section{Introduction}
Liquid fragmentation is the transformation of a compact volume into drops.
The simplest example is the capillary-driven breakup of a slender cylindrical structure
\cite{rutland1971non} at approximately regular intervals driven via the growth of long wavelength perturbations \cite{rayleigh1879a,rayleigh1879b,plateau1849}.
In more general atomization problems, the initial liquid mass deforms transitioning to sheets \citep{bremond, lydia_3},
where the inertial expansion opposed by the capillary deceleration of the edges results in the formation of liquid rims, the subsequent destabilization of which leads to drops. A similar process develops when a perforation occurs in the liquid sheets, then the rapid capillary-driven expansion of holes
\cite{hole_drop, hole_sheet} forms an interconnected set of filaments, which eventually break into drops.  
The evolution of these topological changes are also affected by shear stresses \cite{lasheras,ling}, introducing the effects of Kelvin-Helmholtz \cite{khi} instabilities that induce many of the aforementioned transitions. Given the turbulent regime for atomization, the evolution of liquid structures is chaotic and strongly dependent on initial conditions.
The only common feature that unites these seemingly disparate
fragmentation processes is that the topological stage leading to drop formation is constituted by cylindrical thread-like structures, called ligaments or filaments.

The size of drops resulting from the breakup of ligaments governs 
physical mechanisms underlying a broad range of natural processes and industrial applications. 
These processes include the exchange of heat and mass transfer at 
the ocean-atmosphere interface \cite{deike,seinfeld1998air},
mixing/separation in metallurgical applications \cite{johansen1988fluid,metal},
pesticide dispersal and irrigation in industrial agriculture \cite{bonn, agri_1,agri_2},
and ever so important, pathogen transmission driven by violent respiratory events \cite{lydia_1,lydia_covid}, among many other examples.
Therefore, the development of quantitative models geared towards statistical predictions of the size and velocity
of drops has drawn considerable scientific interest \cite{vill_latest} over the recent decades.

Several experimental and numerical investigations of drop size statistics  
have led to the popularization of three distinct classes of probability density functions,
namely the Log-normal, Gamma and Poisson distributions, as outlined in the review by Villermaux \cite{vill_1}.
In addition, distributions such as the Gaussian \cite{lydia_3}, 
Weibull \cite{weibull}, Exponential \cite{exponential} and 
Beta \cite{beta} have also received significant attention.
Regarding the interpretation of the underlying physical mechanisms, 
the Log-normal model \cite{log_normal} implies a sequential cascade of breakups
(analogous to the Kolmogorov \cite{kolmogorov} energy cascade in fluid turbulence),
the Gamma family \cite{vill_2} considers the competing effects of fragmentation and cohesion,
and the Poisson model \cite{poisson} entails instantaneous and random splitting 
of a volume into smaller fragments. 
These models have been used in a diverse range of fragmentation scenarios 
to varying degrees of predictive success, however, there is a general 
lack of consensus regarding their generalization.
This is primarily due to the fact that the initial liquid structures 
follow markedly different dynamical trajectories towards drop formation, 
rendering certain models incompatible with the actual physical mechanism at play
(refer to \cite{lydia_3} for a discussion).

\subsection{Modes of ligament breakup}

The topological change from the threadlike ligaments to the
(approximately) spherical geometry of drops can proceed along different paths,
depending on the relative importance of viscosity and surface tension, the aspect-ratio, 
and the strength of the initial perturbation \cite{schulkes1996contraction, notz2004dynamics, lohse_asp, wang2019fate}.
Extremely viscous ligaments are stable against capillary-driven disintegration \cite{castrejon2012breakup}. 
For intermediate viscosities, the ligament ruptures at several locations along its length primarily 
due to the Rayleigh-Plateau instability \cite{lohse_asp}. In low-viscosity regimes, the ligament 
might also fragment from one of its free ends, referred to as the end-pinching mode \cite{schulkes1996contraction,end_pinch_2}.  
Additionally, if the ligament is free at both ends and not slender enough (small aspect-ratios),
the capillary retraction might dominate and contract the entire volume into a single drop \citep{stone_1}.
Thus, despite the richness of end-pinching dynamics and complete contraction, 
the resulting drop sizes are extensively documented and well described by 
robust scaling laws \cite{schulkes1996contraction, gordillo2010generation}. 
This turns our attention solely towards the drops formed due to breakups along the ligament length.

\CORR{The breakup mechanism of liquid threads into droplets is fundamentally self-similar \cite{eggers2008physics,eggers1995theory}, corresponding to finite-time singularities in the Navier-Stokes equations. This feature means that the pinching process itself is largely independent of the initial conditions of the ligament. However, the internal liquid dynamics within the ligament are highly sensitive to these initial conditions due to the inherent non-linearities in the governing equations. The distribution of liquid volumes just before the thread ruptures is directly related to the volumes of the resulting droplets. Consequently, precise quantitative control over the initial conditions of the ligament is crucial for understanding the polydispersity in the sizes of the droplets formed.}{The mechanism of the liquid-thread rupture leading to drop formation
is essentially self-similar \cite{eggers2008physics,eggers1995theory}, 
corresponding to finite-time singularities of the Navier-Stokes equations. 
Although this universality renders the pinching process insensitive
to the initial conditions of the ligament, the liquid rearrangements 
within the ligament bulk are sensitive to the initial conditions,
owing to the inherent non-linearities in the governing equations. 
The final arrangement of liquid volumes just prior to the rupture of the liquid-thread
directly correlates to the volume contained in the drops formed.
Therefore, having precise quantitative control over the ligament 
initial conditions is of paramount importance in order to understand 
the polydispersity in the resulting drop sizes.}

\subsection{Our computational framework}

\CORR{In this context, the main goal}{Towards this objective, the central theme} of this study is the design and conception of ``numerical'' experiments, that lend themselves to accurate and repeatable
specifications of the initial conditions of the ligaments in question.
Generally in physical experiments, obtaining ligaments conforming \textit{exactly}
to a specified geometrical shape and velocity field is extremely challenging.
Thus, one often has to employ \textit{a posteriori} correlations between the observed dispersion 
in the final drop sizes and the ``qualitative'' descriptions of initial conditions. 
In contrast, our present numerical framework allows us to obtain reproducible drop size 
distributions, which are purely outcomes of the mathematical model (Navier Stokes with surface tension),
subject to a chosen set of parameters, initial and boundary conditions. 
Furthermore, most of the reported drop size distributions in experiments 
incorporate significant uncertainties, owing to small sample sizes. 
In our case, we are able to precisely control the degree of uncertainty in our 
eventual distributions, as the rapid calculation times enables us to generate large statistical samples. 

\section{Methodology}

\subsection{Mathematical Model}
We use the one-fluid formulation for our system of governing equations,
thus solving the incompressible Navier-Stokes equations throughout the whole
domain, including regions of variable density and viscosity which itself depend on
the explicit location of the interface separating the two fluids \cite{zaleskibook}. 
The interface is modeled as having an infinitesimal thickness at the macroscopic scales under consideration.
The temporal evolution of the interface is tracked by using an advection equation for the 
phase-characteristic function, which is essentially a Heaviside function that distinguishes the individual phases. 
The density and viscosity at each spatial location are expressed as linear functions of the phase-characteristic function.

\begin{figure*}[t]
	\includegraphics[width=0.9\linewidth]{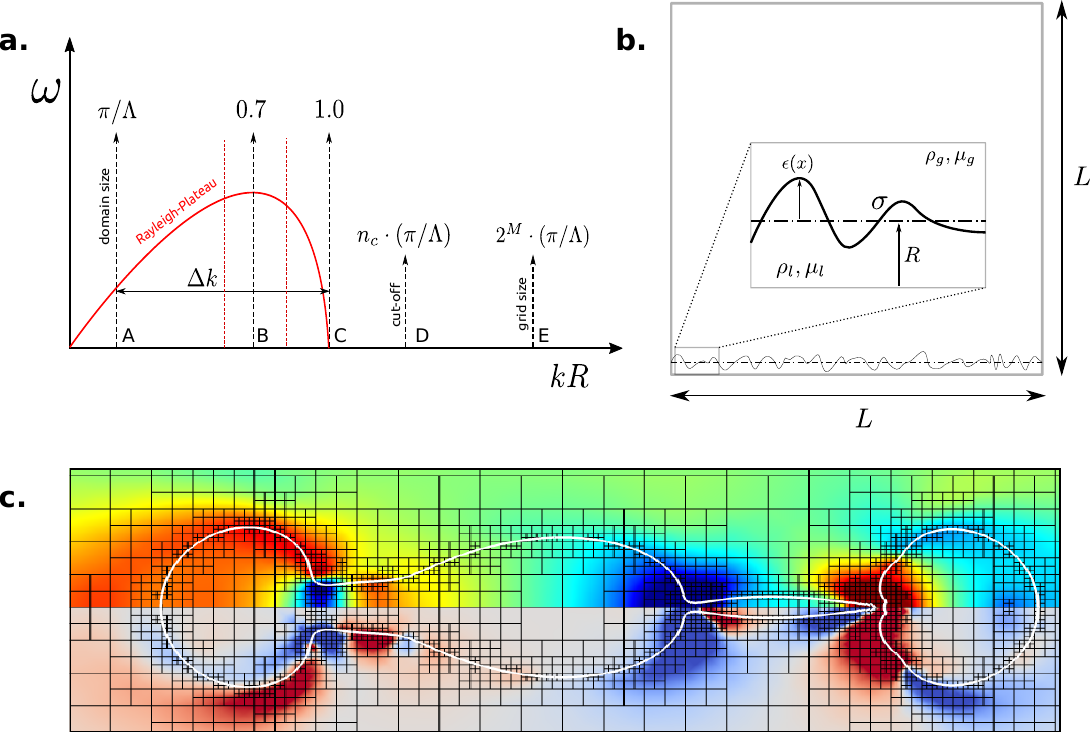}%
	\caption{\label{fig:setup}\textbf{a.} Variation of the linearized growth rate ($\omega$) 
		corresponding to the viscous Rayleigh-Plateau (RP) instability as a function of nondimensional wavenumber $kR$ \cite{weber1931}.
		In our setup, $n_c$ discrete wavelengths are excited as part of the initial condition,
		which fall within the vertical lines $A$ and $D$. 
		Only a certain number of these $n_c$ discrete modes are unstable ($\omega > 0$)
		with respect to the RP instability (red curve, between vertical lines $A$ and $C$). 
		The vertical line $B$ represents the approximate value of $kR$ for which we get the optimal growth rate. 
		\textbf{b.} Schematic of the computational setup.
		An infinitely long and axisymmetric corrugated ligament
		of mean radius $R$ is placed along a side of a square domain of size $L$.
		The bottom side of the box acts as the axis of symmetry, 
		while spatial periodicity is imposed along the horizontal
		direction. Inset : A close up view of the corrugated profile
		of the ligament, where the local radius is defined as
		the sum of the unperturbed (mean) radius $R$ and the local
		perturbation $\epsilon(x)$. The material properties of the
		liquid and gas phases are denoted with the subscripts $l$ and $g$
		respectively, which in our case corresponds to an air-water system
		with the surface tension coefficient $\sigma$.
		\textbf{c.} Dynamically adapted octree meshes near the interface, refined based on
		limiting second gradients of the volume fraction and velocity fields. 
		The interface is represented by the white contours, the colormap \CORR{on the top half is
			based on the axial velocity component, whereas the one on the bottom
			corresponds to that of vorticity.}{the colormap on the left half is based on the axial velocity component, whereas the one on the right corresponds to that of vorticity. }
		The colors red and blue correspond to
		the higher and lower end values respectively, in case of both colormaps.
	}
\end{figure*}

\subsection{Numerical Methods}
We use the free scientific computing toolbox 
Basilisk \cite{basilisk,van2018,pop2015}, which couples finite-volume 
discretization with adaptive octree meshes (see Fig. \ref{fig:setup}c)
in order to solve our governing partial differential equations. 
The interface evolution is tracked using a Volume-of-Fluid (VOF) method \cite{gueyffier, popinet_gerris}, 
coupled with a robust and accurate implementation of height-function based interface curvature computation \cite{popinet2009accurate}. 
The capillary forces are modeled as source terms in the 
Navier-Stokes equations using the continuum surface-force \cite{csf} (CSF) method.

In the present context of ligament destabilization,  
the trajectory of the system towards drop formation is 
governed by non-linear interactions between
capillary waves, remnants of the internal flow,
acceleration of the liquid into the surrounding medium,
localized vorticity production at the interface,
as well as viscous dissipation in the bulk.
In order to accurately reproduce the aforementioned multiscale phenomena
and ensure sufficient spatio-temporal resolution in the vicinity of breakups and coalescence,
the dynamically adaptive octree meshes (Fig. \ref{fig:setup}c.) are absolutely essential
in order to carry out computationally efficient simulations. 
The accuracy and performance of Basilisk has been well documented and extensively validated 
for a variety of complex interfacial flows such as breaking waves \cite{basilisk_1,basilisk_3, wouter}, 
bursting bubbles \cite{basilisk_2,basilisk_5}, drop splashes \cite{basilisk_4}, amongst many others.

\subsection{Computational Setup}
We conduct direct numerical simulations of air-water systems consisting of
slender ligaments with spatial periodicity along the ligament axis.
We use an axisymmetric framework that excludes all 
azimuthal variations in the shape of the ligament and subsequently formed drops.
Fig. \ref{fig:setup}b illustrates the schematic of the
computational setup, where the domain is a square of side $L$.
The bottom side of the box acts as the axis of symmetry for
the corrugated ligament (detailed view in the inset of Fig. \ref{fig:setup}b),
which has an unperturbed (mean) radius $R$.
The radial profile $R(x)$ along the ligament axis can be written as 
$R(x) = R + \epsilon(x)$, where \CORR{$\epsilon(x)$ is considered to be a perturbation following a normal distribution with a mean value of 0 and variance $\varepsilon^{2}_{0}$}{$\epsilon(x) \sim \mathcal{N}\left(0,\varepsilon^{2}_{0} \right) $}. 
Periodic boundary conditions are imposed for the primary variables
on the left and right faces of the domain.
Symmetry boundary conditions are imposed on the bottom side,
with the impenetrable free-slip condition applied to the top side.

\subsubsection{Random Surface Generation}
\CORR{The random surfaces of our spatially periodic ligaments are generated using a white noise signal, which is produced by a robust random number generator \cite{rng}. This signal is then filtered to retain only the longest $n_c = 25$ wavelengths, given that only these are relevant for hydrodynamic instabilities. resulting in the final radial profile of the ligament with a variance of $\varepsilon_0^2$. The surface profile of each individual ligament in the ensemble is uniquely determined by the seed of the random number generator \cite{rng}. This method allows us to create an ensemble of ligaments with random but unique surface profiles by varying the seed values.

For infinitely long ligaments, only perturbations with wavelengths longer than the ligament circumference are unstable to the Rayleigh-Plateau type capillary instability \cite{rayleigh1879a,rayleigh1879b}. Due to the discrete nature of numerical simulations, we can initially excite only a finite number of discrete modes that lie within the unstable spectrum (see Fig. \ref{fig:setup}a). The number of these unstable discrete modes is proportional to the ligament aspect-ratio $\Lambda = L/W$ ($\Delta k \sim \CORR{\pi/\Lambda}{\Lambda / \pi}$). In our simulations, we have 15 discrete unstable modes, including several close to the optimal Rayleigh-Plateau wavelength.}{The random surfaces of our spatially periodic ligaments 
are constructed by taking a white noise signal,  
(using a robust random number generator \cite{rng})
which is subsequently filtered (keeping only longest $n_c = 25$ wavelengths) 
in order to generate the final radial profile of the ligament with variance $\varepsilon_0^2$.
The exact surface profile of an individual ligament in the ensemble is precisely and uniquely
determined by the ``seed'' (state) of the random number generator \cite{rng}, 
thus allowing us to create an ensemble of such random but unique surface 
profiles by mapping each profile to unique values of the seed. 
In the case of infinitely long ligaments, only perturbations with 
wavelengths larger than the ligament circumference are unstable to 
the Rayleigh-Plateau \cite{rayleigh1879a,rayleigh1879b} type capillary instability. 
Owing to the discrete nature of numerical simulations, we are only able
to initially excite a finite and small number of discrete modes that fall within 
the unstable part of the spectrum (see Fig. \ref{fig:setup}a).
The number of such unstable discrete modes varies linearly
with the ligament aspect-ratio $\Lambda = L/W$ ($\Delta k \sim \Lambda / \pi$), therefore, 
in our case we have $15$ discrete unstable modes, including a few close 
to the optimal Rayleigh-Plateau wavelength.}

\subsubsection{Regime of Interest}
In order to isolate the influence of
initial geometrical shape on the
subsequent dynamics and drops formed, we exclude inertial forces (axial stretching rate) 
in our initial conditions. 
The mean radius $R$ of the ligament is the characteristic 
length scale of the problem.
As we are dealing with air-water systems (20 degrees Celsius), the density and viscosity ratios 
are given as $\rho_l / \rho_g \simeq 830$ and $\mu_l / \mu_g \simeq 45$ respectively. 
Thus, our system is characterized by the Ohnesorge number which is defined as 
\begin{equation}
	\textrm{Oh} = \mu / \sqrt{\rho \sigma R} \,.
\end{equation}
The Ohnesorge number is simply the square-root of the ratio
of the viscous-capillary length scale ($l_{\mu} = \mu^2 / \rho \sigma $) 
with the characteristic length scale of the problem ($R$). 
Although the configuration initially has no kinetic energy,
a part of the surface potential is immediately converted into 
liquid inertia as soon as the system is released from its static initial conditions.
The geometrical shape of any individual ligament in our ensemble 
is characterized by a mean corrugation amplitude $\eta = \varepsilon_0 / R$, and
aspect-ratio $\Lambda = L / W$, where $W=2R$ denotes the mean width of the ligament. 
The volume of the corrugated ligament per unit spatial period ($L$)
is controlled by $\Lambda$, which also acts as the
theoretical upper bound to the drop size.
Additionally, we rescale physical time with the capillary time scale such that
$T = t  / t_{\sigma}$, where $t_{\sigma} = \left(\rho R^3 / \sigma \right)^{-1/2}$.
The material properties used in our adimensional parameters 
($\rho$, $\mu$) correspond to the liquid phase i.e. water. 
In the present study, we focus our attention on \textit{weakly}
perturbed ($\eta \simeq 0.08$) and sufficiently slender ligaments ($\Lambda \simeq 50$)
\CORR{with $\textrm{Oh} \simeq 10^{-2}$, which correspond to water ligaments of a diameter close to a $100$ microns, representing the dynamics of the experiments on \cite{wang2019fate,lydia_3} within a relevant range of dimensionless parameters\cite{pal2020investigation,zaleski2021droplet}.}{at the characteristic length scale of $100$ microns ($\textrm{Oh} \simeq 10^{-2}$)}.

\begin{figure*}[t]
	\centering
	\includegraphics[width=0.9\linewidth]{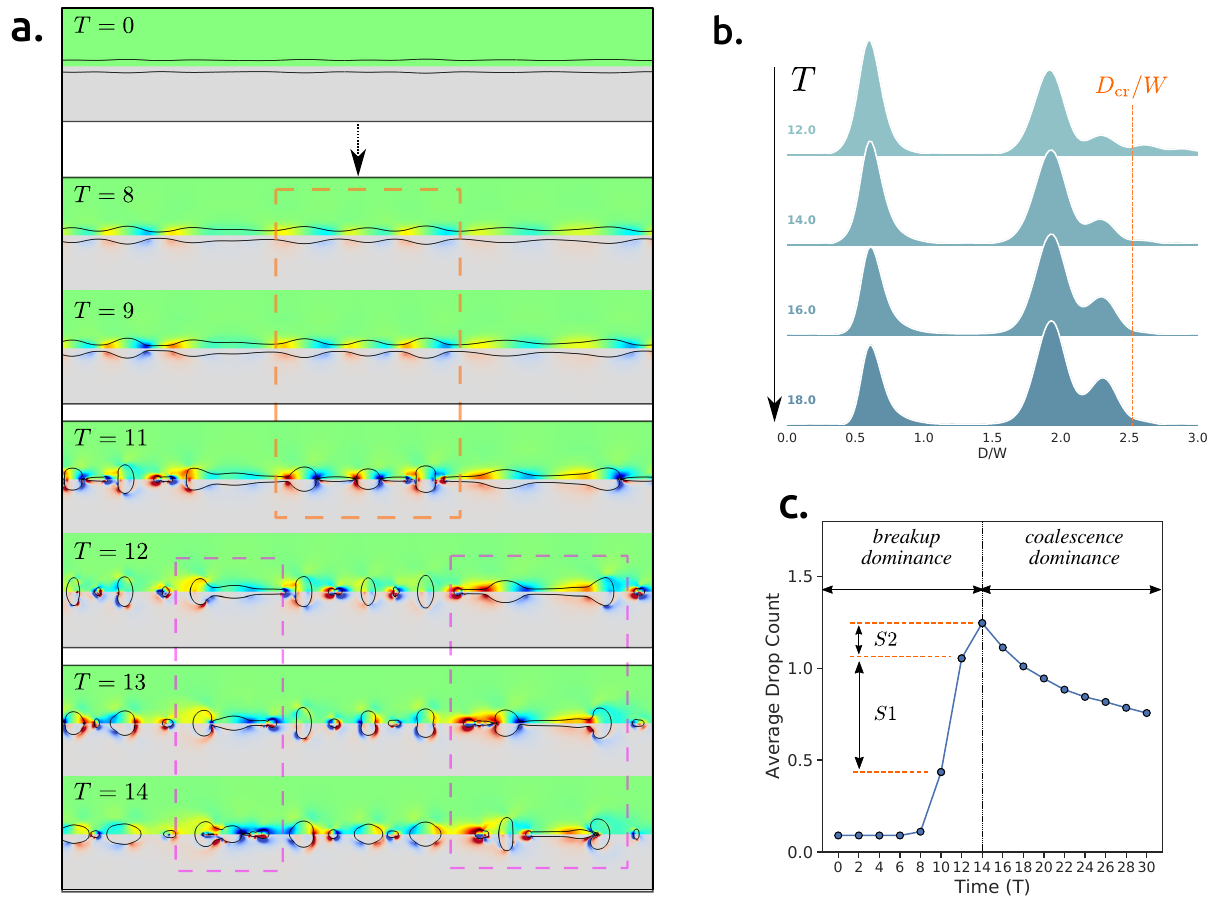}
	\caption{
		\textbf{a.} Destabilization of a \CORR{typical ligament through the breakup stages}{ligament randomly selected from our ensemble, demonstrating the different breakup stages}. The interface is represented by black contours.
		Colormaps of the top and bottom halves in each plot represent the axial velocity and the vorticity magnitude respectively.
		In snapshots $T=8,9,11$, we observe the formation of drops (orange dashed box) 
		corresponding to the optimally perturbed wavelength of the Rayleigh-Plateau instability.
		This leads to the first stage of breakups ($\textbf{S1}$), where the ligament disintegrates into 
		primary, satellite and secondary drops, along with some elongated structures. 
		\CORR{The latter (purple dashed boxes), disintegrate into smaller sizes during the second stage of breakups ($\textbf{S2}$).}{
		Subsequently, we enter the second stage of breakups ($\textbf{S2}$), where the elongated structures 
		themselves disintegrate into smaller sizes, within purple dashed boxes.
		}
		\textbf{b.} Temporal evolution of the probability density functions of \CORR{dimensionless drop size ($D/W$).}{drop size, where drop size is expressed as diameter rescaled by the initial ligament width $W$.}
		Across time, the distribution peaks reveal three stable drop sizes, namely \CORR{satellite ($D/W \simeq 0.6$), primary ($D/W \simeq 1.9$), and secondary ($D/W \simeq 2.3$) drops.}{the satellite drops ($D/W \simeq 0.6$), primary drops ($D/W \simeq 1.9$) and secondary drops ($D/W \simeq 2.3$).}
		The number of satellite drops decreases with time
		due to coalescence with adjacent larger drops. \CORR{The number of primary and secondary drops increases with time}{The secondary drops grow in number as time progresses,} due to the continuous breakup of the elongated structures with aspect-ratios above the critical threshold ($\Lambda_{\textrm{cr}} / W \sim (D_{\textrm{cr}}/W)^3 $).
  \textbf{c.}  The average number of drops \CORR{in the ensemble as a }{generated through the disintegration of the ligaments in our ensemble, mapped as a }function of time. 
		\CORR{Before $T=6$, breakup is rare.}{A limited number of breakups occur before $T=6$.}
		Starting from $T=8$, breakup events occurring on much faster timescales, 
		leading to a peak in number of drops at $T=14$. 
		Beyond $T=14$, \CORR{coalescence dominates, leading to a slower decrease of the average drop count}{the number of breakup events is significantly less than coalescence, thus leading to the average drop count decreasing over a slower timescale.} 
	}
	\label{fig:drops_vs_time}
\end{figure*}

\section{Results}
The process of drop formation via ligament breakup is deterministic, 
therefore it is completely characterized by the initial geometrical shape of the ligament.
Stochasticity is introduced by creating an ensemble of such corrugated ligaments,
where each individual case has a random and unique surface. 
The statistical properties of the 
corrugated shape are identical across all ligaments in the ensemble.
This key step allows us to incorporate the effects of the myriad underlying
processes that determine the exact ligament shape in realistic fragmentation scenarios, 
that too in a quantitatively precise and reproducible manner. 

\subsection{Statistics of Drop Formation}
In Fig. \ref{fig:drops_vs_time}a, we illustrate the different stages involved in
the breakup of an individual ligament into drops, where the ligament is 
randomly selected from our ensemble of size $10000$.  
Linear theory based on the Rayleigh breakup \cite{rayleigh1879a,rayleigh1879b} of infinitely
long liquid cylinders in a quiescent medium predicts the
initial destabilization phase (panels $T=8$, $T=9$ of Fig. \ref{fig:drops_vs_time}a) 
proceeding via exponential growth of the different (unstable) discrete 
frequencies that constitute the initial surface perturbation. 
Beyond this linear growth phase, non-linearities rapidly kick in 
near the breakup zones \cite{sat_1,sat_2, rutland1971non}, 
eventually resulting in the formation of ``main'' and significantly 
smaller ``satellite'' droplets , as observed in panels $T=11$, $T=12$ of Fig. \ref{fig:drops_vs_time}a. 
In our study, we refer to this as the first stage of breakups ($\textbf{S1}$),
where we find a set of ``primary'' and ``satellite'' drops (orange dashed box in Fig. \ref{fig:drops_vs_time}a), 
along with a collection of strongly deformed elongated structures (purple dashed box in Fig. \ref{fig:drops_vs_time}a)
which themselves resemble small aspect-ratio ligaments. 
This stage is immediately followed by the second stage of breakups ($\textbf{S2}$),
in which the elongated structures break down into smaller fragments,
while the previously formed primary and satellite drops remain stable. 

The number of drops in our ensemble is measured using \textit{average drop count}, 
\CORR{defined as the ratio between the total number of drops in the ensemble
to the total length of the ligament ensemble, measured in critical wavelength ($\lambda_{\text{RP}} = 2\pi / k_{\text{RP}}\simeq 9R$), corresponding to the maximum growth rate of the viscous Rayleigh-Plateau instability \cite{weber1931,rayleigh1879a,rayleigh1879b}.}{defined as the ratio between the total number of drops in the ensemble
and the entire extension of the ligament ensemble, measured in \textit{characteristic length}. The characteristic length is chosen as the wavelength ($\lambda_{\text{RP}} = 2\pi / k_{\text{RP}}\simeq 9R$)  corresponding to the optimal growth rate of the viscous Rayleigh-Plateau instability \cite{weber1931,rayleigh1879a,rayleigh1879b}.}
In Fig. \ref{fig:drops_vs_time}c, we plot the temporal variation of \textit{average drop count}.
The slope of the graph is determined by the competition between breakup and coalescence events, 
thus delineating the two distinct stages of breakup ($\textbf{S1}$ and $\textbf{S2}$), as well 
as the dominance of coalescence events beyond $T=14$, leading to a slow decrease in the number of drops.  

Coming to the statistics of drop sizes, in Fig. \ref{fig:drops_vs_time}\CORR{b}{a}, 
we show the probability density functions (PDF)
corresponding to drop size distributions as a function of time. 
The drop diameters are re-scaled by the initial width ($W$) of the ligaments.
One can clearly observe the presence and persistence of three distinct peaks
in the size distribution for all instants of time shown. 
These stable peaks correspond to drop sizes given by 
$D/W \simeq 0.6$ for the satellite drops, $D/W \simeq 1.9$ for the primary drops,
and $D/W \simeq 2.3$ for what we refer to as ``secondary'' drops.

Assuming that drops are formed by encapsulating the volume of liquid contained within one optimal
wavelength ($2\pi / k_{\text{RP}}$), we can compute the diameter $D_{\text{RP}}$ as

\begin{align}
	\frac{\pi}{6} D_{\text{RP}}^3 &= \frac{\pi}{4} W^2 \left( 2\pi / k_{\text{RP}}\right)
	\implies D_{\text{RP}} / W \simeq 1.89  \,. \label{lin}
\end{align}

As we can observe in Fig. \ref{fig:drops_vs_time}\CORR{b}{c}, the statistical estimate of 
our primary drop size (values distributed around $D/W \simeq 1.9$) across time is in excellent agreement 
with the predictions \eqref{lin} of linearized stability theory. 

The typical size of satellite drops has a strong dependence on 
the initial conditions, as meticulously documented in the seminal 
work of Ashgriz \& Mashayek \cite{ashgriz_sat} concerning the capillary breakup of jets. 
In that study, the authors report a monotonic decrease in the satellite drop size 
as one increases the initial perturbation strength (Fig. 12 in \cite{ashgriz_sat}). 
At the limit of vanishing perturbation strength (matching our initial conditions), 
Ashgriz \& Mashayek obtain a satellite drop size of $D/W \simeq 0.6$, 
which matches quite well with the statistical observations of our satellite drop size (Fig. \ref{fig:drops_vs_time}b). 

Immediately after the first set of breakups ($\textbf{S1}$), there are plenty of
elongated structures with free ends (including our ``secondary'' drops), which might be subject to the end-pinching mechanism. 
Several numerical, experimental and scaling analyses in existing literature 
(Shulkes \cite{schulkes1996contraction}, Gordillo \& Gekle \cite{gordillo2010generation}
,Wang \& Bourouiba \cite{lydia_3}) have established that the size of drops 
generated via the end-pinching mechanism are deterministically characterised 
by the width of the ligament of origin, given by a near constant value of $D/W \simeq 1.5$ 
(although with an extremely weak dependence on inertial stretching rate).
Therefore, the absence of any peak in our drop size statistics (Fig. \ref{fig:drops_vs_time}c)
after $T=12$ (beyond $\textbf{S1}$) around the value $D/W \simeq 1.5$ is a striking 
observation, asserting that negligible breakups occur via the end-pinching mode. 
Further investigations must be conducted in order to establish the exact cause  
of this absence.

\begin{figure*}[t]
	\centering
	\includegraphics[width = 0.9\linewidth]{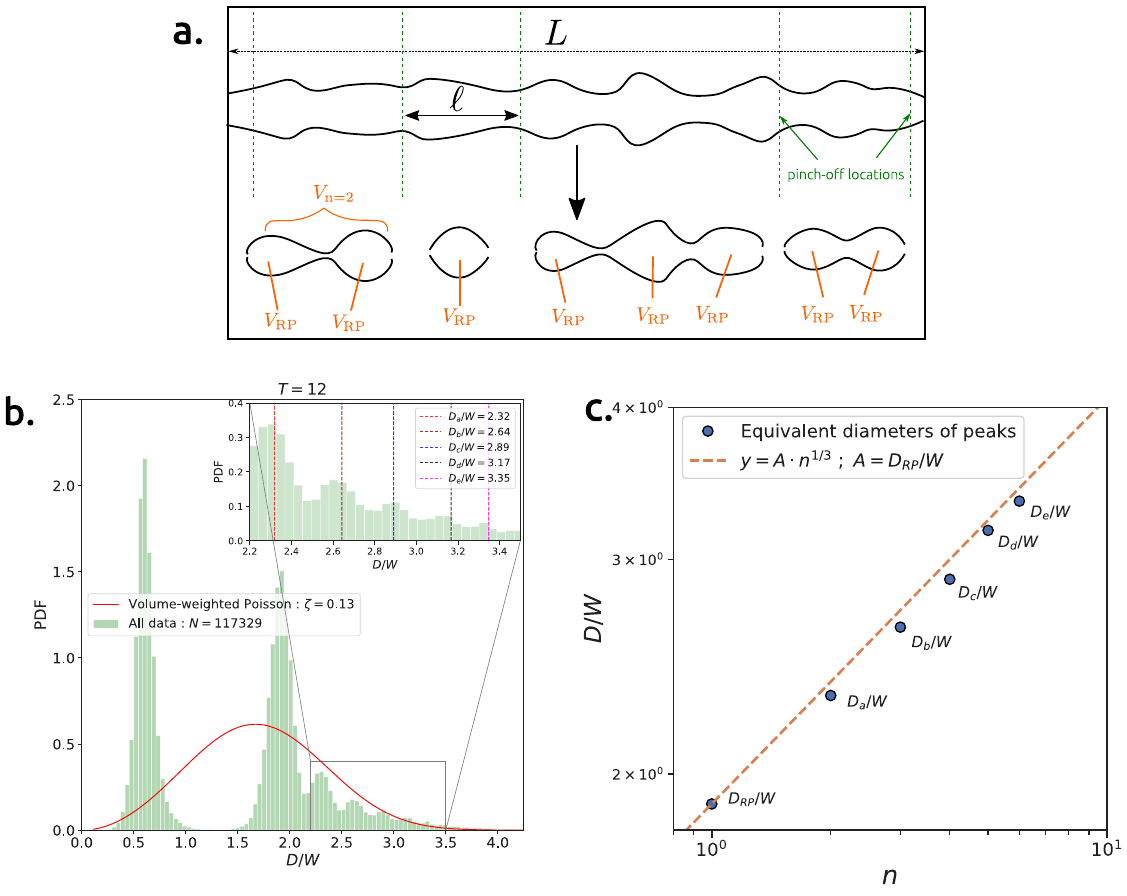}
	\caption{\textbf{a.} Representation of the first stage of breakup dynamics ($\textbf{S1}$).
		The green vertical lines on the intact ligament (top figure) denote the possible pinch-off
		locations, \CORR{spaced by an $\ell$ distance}{and $\ell$ is the random variable describing the exponential spacing (Poisson model) between successive pinch-off events}, along the length $L$ of the ligament. 
		\CORR{After pinch-off}{Once the pinch-offs occur}, the volume $V_n$ of the ``elongated drops'' 
		can be modeled as \CORR{multiple of $V_{\rm RP}$, which corresponds to one unit of the optimally perturbed viscous Rayleigh-Plateau instability wavelength}{a linear combination of smaller characteristic volumes, where $V_{\rm RP}$ belongs to normal distributions centered around the expected value of volume contained under one unit of the optimally perturbed viscous Rayleigh-Plateau instability}.
		\textbf{b.} Probability density function of the drop size at $T=12$, displaying the peaks 
		corresponding to the satellite and primary drops, as well as the typical sizes of
		the elongated drop-like structures in the distribution tail. 
		The volume-weighted Poisson distribution \eqref{eq:vwp} 
		is plotted using a pinch-off rate ($\zeta = 0.13$) determined 
		by the average number of drops formed per ligament. 
		Inset: Zoom-up on the peaks representing the typical sizes of the elongated structures.
		\textbf{c.} The predictions of our simplified model (orange dashed line) for the typical sizes of
		the elongated structures, plotted alongside the statistical observations of the drop sizes
		(blue circles) that constitute the peaks within the tail of our distribution at $T=12$. 
		Assuming that the elongated structures are generated by encapsulating integer multiples of 
		the characteristic volume $V_{\rm RP}$, the equivalent diameters should scale according to $D_n / W \sim n^{1/3}$, 
		where $n$ is the number of characteristic units of $V_{\rm RP}$. 
	}
	\label{fig:pdf}
\end{figure*}

\subsection{First Stage of Breakups ($\textbf{S1}$)}
We take a closer look at the probability of the large drop sizes
immediately after the first set of breakups. 
We start with a simple model for the ligament pinching-off
at several locations, with 
\CORR{the assumption of a small, uniform and 
independent probability of the ligament pinching-off in each small length element ${\rm d}x$.
Therefore
the spacing $\ell$ between
any two pinch-off locations (see Fig. \ref{fig:pdf}\CORR{a}{b}) follows an exponential probability distribution
\begin{equation}
	P_1(\ell) = \zeta \, \textrm{exp}(-\zeta \ell) ,
\end{equation}
where $\zeta$ is the average number of pinch-offs occurring over a unit length. In other words the probability that the pinched-off length is between $\ell$ and $\ell + {\rm d}\ell$ is $P_1(\ell) {\rm d}\ell$.
It is then easy to show \cite{yates2014probability} that the  number $n$ of pinch-offs over a length $x$ 
follows the Poisson distribution
\begin{equation}
    P_P(n;x) = \frac{(\zeta x)^n}{n!} \exp ( - \zeta x ). \label{eq:lwp}
\end{equation}
The volume of the drop formed by encapsulating the volume between two successive pinch-off locations separated by a distance $\ell$
is clearly $V = \pi W^2 \ell / 4$ and the diameter of a spherical droplet with that volume is 
$d = [(3/2) W^2 \ell]^{1/3}$.
Using the change of variable relation for probability distributions 
$P_1(\ell) {\rm d}\ell = P_d(d) {\rm d} d$
we obtain the expression for the PDF of the diameters as
\begin{equation}
	P_d(d) = P_1(\ell) \frac{{\rm d}\ell }{ {\rm d} d } = \frac{2 \zeta d^2}{ W^2} \textrm{exp}[- 2\zeta d^3 / (3 W^2) ] . \label{eq:vwp}
\end{equation}
Thus the diameter distribution is identical to a Poisson distribution with $n=2$ and we  refer to it
as the ``volume-weighted'' Poisson distribution arising from an exponential distribution of
pinch-off intervals.}
{the assumptions 
(i) only one pinch-off occurs in an 
infinitesimally small interval $dl$.(ii) a small, uniform and 
independent probability of the ligament pinching-off in each $dl$.
Therefore, the total number of pinch-offs over the entire length $L$ 
follows a Poisson distribution, implying that the spacing $\mathcal{L}$ between
any two pinch-off locations (see Fig. \ref{fig:pdf}b) follows an exponential distribution, with 
the probability density function given by 
\begin{equation}
	f_{\mathcal{L}}(x) = \zeta \cdot \textrm{exp}(-\zeta x) \,,
\end{equation}
where $\mathcal{L}$ is the random variable for the exponential spacing,
and $\zeta$ being the expected value of the number of pinch-offs occurring over length $L$. 
We set $\mathcal{D} \equiv D/W $ as the random variable representing 
the size of the drop formed by encapsulating the volume between two successive pinch-off locations.
Using the relation $\mathcal{D} = c \mathcal{L}^{1/3}$ ($c$ is a constant), 
we obtain the expression for the PDF of random variable $\mathcal{D}$ as 
\begin{equation}
	f_{\mathcal{D}}(x) = \frac{3 \zeta x^2}{c} \textrm{exp}(-\zeta x^3 / c ) \,, \label{vwp}
\end{equation}
which we refer to as the ``volume-weighted'' exponential (or Poisson) distribution.}
The pinch-off rate \CORR{for our numerical experiment is determined by first observing the number of 
drops formed}{is determined by first calculating the rate of 
drops formed }: $117,329 \, \text{drops} / 10,000 \,  \text{ligaments} \approx 12$ drops per ligament. 
Since the average number of pinch-offs is equal to \textit{one more} than the average number of drops,
we obtain $\zeta \simeq 0.13$ as there are 100 units of length per ligament.
In Fig. \ref{fig:pdf}b, we plot the PDF of the drop size distribution at $T=12$,
as well the volume-weighted exponential PDF using $\zeta = 0.13$ (no free parameters).
We observe that the tail of the distribution matches the predictions of the 
volume-weighted exponential model of Eq. (\ref{eq:vwp}) quite satisfactorily,
even though it cannot capture the probabilities of the primary and satellite drops. 

We observe that the tail of the distribution at $T=12$ contains small peaks 
(see Fig. \ref{fig:drops_vs_time}a at $T=12$, Fig. \ref{fig:pdf}b), which corresponds
to some \textit{typical} sizes of the elongated structures, and we seek
a simplified model to predict the size of such drops.
As demonstrated in Fig. \ref{fig:pdf}a, each ``elongated drop'', formed after the Poisson-like pinch-off events, is assumed to be a connected set of smaller \textit{characteristic} volumes.
\CORR{We consider, following the $V$ definition given between Eqs. (\ref{eq:lwp}) and (\ref{eq:vwp}), a characteristic volume $V_{\rm RP} = \frac{\pi}{4} W^2 \left( 2\pi / k_{\text{RP}}\right)$}{Opting for a further simplification, we only consider one characteristic volume with value $V_{\rm RP} = \frac{\pi}{4} W^2 \left( 2\pi / k_{\text{RP}}\right)$} to model any given elongated structure as composed of integer multiples of the volume encapsulated under optimal viscous Rayleigh-Plateau wavelength. 

Thus, given $V_n = n \cdot V_{\rm RP}$, the diameters 
corresponding to the peaks in the inset of Fig. \ref{fig:pdf}b should simply vary
as $D_n / W = A \cdot n^{1/3}$, where $A = D_{\rm RP} / W $ is the Rayleigh-Plateau optimal drop size.
In Fig. \ref{fig:pdf}c, we plot the predictions of this simple model against the 
statistically observed values of the peaks present in the distribution tail at $T=12$ (inset Fig. \ref{fig:pdf}b) .
The close agreement of our model with the statistical observations strongly suggests that the volume of each of the elongated structures
(sizes larger than the primary drops) is close to an integer multiple of the volume of the typical RP primary drop. The 
number of primary drop volumes within one elongated structure is determined by exponentially distributed  pinch-offs along the  ligament. 
\begin{figure*}[t]
	\centering
	\includegraphics[width=0.9\linewidth]{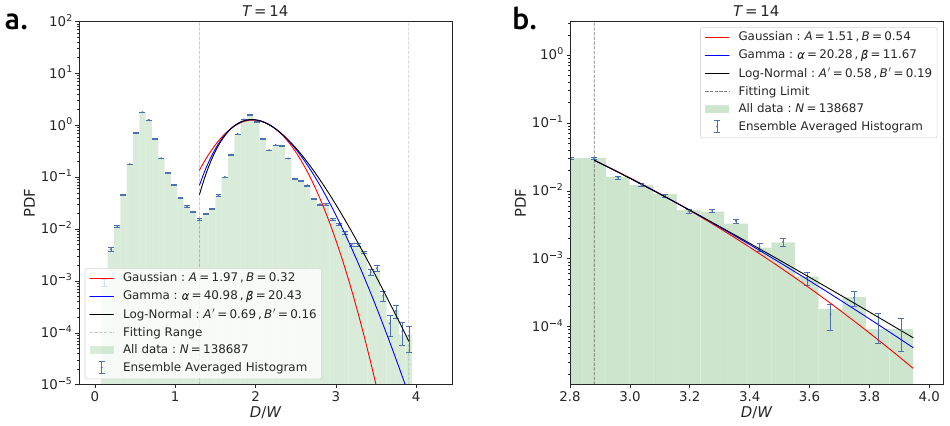}
	\caption{ Drop size distributions at $T=14$, representing the drop ensemble
		immediately at the \CORR{highest}{higest} value of \textit{average drop count}.
		Averaged distributions (blue points with error bars) with $95\%$ confidence intervals are plotted
		on top of the distribution corresponding to the entire ensemble of size $N=138,693$ (green histogram). 
		\textbf{a.} The best fits corresponding to the Gaussian, Log-Normal and Gamma distribution
		functions are plotted within a range (dashed vertical lines) that \textit{includes}
		the peak representing the primary drops.
		We observe that the Log-Normal fit best describes the distribution over the selected range, 
		and differences comparing the Gamma fit appear only near the tail end of the distribution. 
		\textbf{b.} The best fit corresponding to the Gaussian, Log-Normal and Gamma distribution
		are plotted while \textit{excluding} the peak representing the primary drop size. 
		\CORR{For this range of sizes, Log-Normal and Gamma fits are still closer to the data,}{In this case, the Gamma fit appears to best describe the probabilities of the large drop sizes,} but it is difficult to distinguish between each of the three candidate functions. 
	}
	\label{fig:fitting}
\end{figure*}

\subsection{Second Stage of breakups ($\textbf{S2}$)}
We now turn our attention towards the large ($D/W > 1.9$)
``elongated'' drops during the second stage of breakups.
Considering these structures as small aspect-ratio ligaments, 
they can collapse into a single (or two) drop(s) via the 
capillary-driven retraction of both the ends,
or break up into multiple drops along its length through a Rayleigh-Plateau
type instability mechanism.
Driessen et al. \cite{lohse_asp} demonstrate using a 
combination of analytical arguments and numerical
simulations, the existence of a critical aspect-ratio $\Lambda_{\text{cr}}$, below which,
the structure is entirely stable against the Rayleigh-Plateau instability. 
This critical $\Lambda_{\text{cr}}$ is determined by equating the time
taken by the optimal Rayleigh-Plateau perturbation to grow to the ligament radius,
with the time taken by the two ends to retract to half the ligament length.
The expression for $\Lambda_{\text{cr}}$ provided by Driessen et al. \cite{lohse_asp}, but adapted to our problem setup is given as 

\begin{equation}
	\frac{|\textrm{log}(\eta^\prime)|}{t_\sigma \cdot \omega_\textrm{max}} 
	+ \left(6\Lambda_{\text{cr}} \right)^{1/3} - \Lambda_{\text{cr}} = 0 \,,
	\label{eqn:aspect}
\end{equation}

where, $\eta^\prime$ indicates the degree to which 
the surface of the ligament (elongated drop) is perturbed.
The perturbation strength corresponding to our initial condition $\eta$
acts as the lower bound to $\eta^{\prime}$ simply due to the fact 
that the perturbations grow as a function of time. 
The optimal growth rate $\omega_{\textrm{max}}$ is a function of the 
Ohnesorge number, and is calculated from the dispersion relation 
obtained by Weber \cite{weber1931} (Fig. \ref{fig:setup}a) for the capillary instability 
at the low Reynolds limit of the Navier-Stokes equations.
Using a simple root-finding algorithm for the non-linear equation \ref{eqn:aspect},
with $\eta^{'} = \eta$ , we obtain the critical aspect-ratio value 
for our setup as $\Lambda_{\text{cr}} \simeq 11.5$, which is a slight overestimation
due to the fact the our elongated structures are significantly more perturbed than $\eta$.
Computing the equivalent diameter for the volume contained in a ligament of mean width $W$ 
and aspect-ratio $\Lambda_{\text{cr}}$, we get $D_{\text{cr}}/W \simeq 2.5$. 

Revisiting Fig. \ref{fig:drops_vs_time}b, we observe that the 
number (or probability) of drops lying to the right of 
the $D_{\text{cr}}/W$ mark (orange dashed line) decreases 
with time starting from $T=12$ to $T=16$.
In addition, the ``secondary'' peak is the only 
one whose height does not decrease with time, rather, increases steadily with time.
This observation can be explained by the continuous breakup of the 
elongated drops into smaller fragments, till they finally attain aspect-ratios
just below the critical threshold $\Lambda_{\text{cr}} \simeq 11.5$, 
at which point they become immune to any further capillary instability.
Looking at peak representing the ``secondary'' drops, we observe that  
they lie just below the critical threshold $D_{\text{cr}}/W \simeq 2.5$, therefore 
demonstrating a qualitative match between the statistical observations of our simulations and 
the predictions of the ``Driessen model\cite{lohse_asp}'' (Eq. \ref{eqn:aspect}).

Finally, we study the drop size distributions
immediately after the second stage of breakups $\textbf{S2}$ at $T=14$. 
In terms of candidate probability density functions 
for the large drop sizes, we use the three most popular
choices in existing literature, namely, 
the Gaussian , Log-Normal and Gamma distributions
(definitions in the Appendix). 

In Fig. \ref{fig:fitting}, we plot the best fits pertaining to
the aforementioned candidate functions on a log-linear scale, 
within different ranges of interest (vertical dashed lines). 
The histogram bins are ensemble averaged, where the $95\%$ confidence intervals
are computed using a standard bootstrap re-sampling procedure 
(refer to the Appendix).
Fig. \ref{fig:fitting}a demonstrates that by including the peak
representing the primary drops, the distribution is \textit{roughly} 
described by a Log-Normal distribution, where significant differences 
from the Gaussian and Gamma fits mainly appearing near the tail ($D/W > 3$). 
Subsequently, in Fig. \ref{fig:fitting}b we restrict our focus 
to the tail, therefor excluding the primary drop peak. 
We observe the Log-Normal and Gamma fits appears to describe the probabilities of large sizes with similar accuracy, better than the Gaussian. This last fit misses the error bars near the tail, while the other two fall within range. 
It is important to note that the upper limit to the drop size 
is given by the volume of the entire ligament; for our
considerably slender ligaments ($\Lambda \simeq 50$),
the largest drop size is given by $D_{\text{max}}/W \simeq 4.2$.
Thus, even while having converged statistics, sufficiently large samples
and robust error bars, there is a fundamental limitation when it comes 
to distinguishing between the curvatures of our exponential candidate  
functions near the tail region, simply as a consequence of the 
restricted range ($ 1.9 < D/W < 4.2$) of drop sizes.

\section{Conclusions \& Perspectives}

The fragmentation of liquid masses in high-speed flows, such as atomizing jets, breaking waves, 
explosions or liquid impacts, is of utmost practical importance, and of interest for the statistical study of flows. 
Although drop size distributions can be inferred from experiments, 
our high-fidelity numerical approach crucially provides the direct predictions 
of the mathematical model i.e. Navier-Stokes with surface tension. 
Here we explore this distribution for the simplified case of a liquid ligament, where the simplification 
allows us to obtain high-fidelity solutions for ensembles that are so 
large that the statistical error is smaller than in most experiments to date. 
Thus, this study constructs a solution to the distribution problem based directly on the 
conventionally accepted mathematical model, that too in a quantitatively precise, 
statistically robust and reproducible framework. 

Our statistical distributions reveal three stable drop sizes, generated via a sequence of two 
distinct breakup stages. After the 1st stage, the probability of the large sizes are shown to 
follow a parameter-free volume-weighted Poisson distribution, but immediately after the second breeakup 
stage, the large sizes are best described by a two-parameter Log-Normal distribution, although the
Gamma distribution seems to be the best fit for the distribution tail. 
Finally, we also point out that due to the small range of drop sizes involved, 
it is inherently difficult to distinguish between the curvatures of different exponential curves. 

Moving forward, we would like to find a quantitative explanation concerning the 
absence of the end-pinching drop formation mode in our observations. 
In addition, we would like to verify the consistency of our findings across 
a broad span of length scales corresponding to $ 10^{-4}< \rm{Oh} < 1$. 
The essential next steps in our effort towards developing a 
higher fidelity picture of realistic fragmentation scenarios would
involve incorporating additional layers of complexity on top of our 
simplified ligament model, such as a stretching flow, 
turbulent fluctuations in both liquid and gas phases, as well as high shear rates at the interface. 

\begin{acknowledgments}
This work has benefited from access to the HPC resources of CINES under the allocations 2018- A0052B07760 and 2019 - A0072B07760, and the resources of TGCC under the project 2020225418 granted respectively by GENCI and PRACE and by the Flash Covid grant of HPC resources. Support by the ERC ADV grant TRUFLOW and by the Fondation de France for the ANR action Flash Covid is gratefully acknowledged. We thank Lydia Bourouiba for fruitful discussions regarding liquid fragmentation, Gareth McKinley for planting the seed of the idea, and St\'ephane Popinet for the Basilisk platform.
\end{acknowledgments}

\appendix*
\section{}
Our drop population $\mathcal{P}$ at $T=14$ has a size equal to $138,693$.
From $\mathcal{P}$ we draw a random sample of size $10000$, which we denote as $\mathrm{S}_1$.
Repeating this sampling procedure (with replacement) 200 times, we create an ensemble 
of such samples $\mathcal{E}_j = \{\mathcal{S}_1, . . . , \mathcal{S}_{200} \}_j $.
Histograms are generated for all samples in $\mathcal{E}_j$, given a fixed set of binning intervals.
An ensemble averaged histogram for $\mathcal{E}_j$ is obtained by computing the mean
of the corresponding bin heights over all samples $\mathcal{S}_i$,
which are plotted in Fig. \ref{fig:fitting} (blue points with error bars). 
The standard error on the ensemble averaged bin heights is computed using bootstrapping:
(i) the ensembling procedure is repeated to construct $50$ such ensembles ($\{\mathcal{E}_1 , ... , \mathcal{E}_{50} \}$),
(ii) ensemble averaged histograms are computed for each $\mathcal{E}_j$ as previously described,
(iii) the standard deviation of the average bin heights across $\{ \mathcal{E}_1, ..., \mathcal{E}_{50}\}$ gives us the standard error.   
The error bars in Fig. \ref{fig:fitting} represents a range of 4 standard deviations i.e. $95\%$ confidence intervals. 
The probability density functions are defined as

\begin{align*}
        \text{Gaussian : } \, P\left( x ; A , B \right) &=
        \frac{1}{B \sqrt{2\pi}} \textrm{exp}\left[-\frac{1}{2}\left(\frac{x - A}{B}\right)^2\right]  \,,\\
        \text{Log-Normal : } \, P\left( x ; A^{\prime} , B^{\prime} \right) &=
        \frac{1}{x B^{\prime} \sqrt{2\pi}} \textrm{exp}\left[-\frac{1}{2}\left(\frac{\log x - A^{\prime}}{B^{\prime}}\right)^2\right] \,,\\
        \text{Gamma : } \, P\left( x ; \alpha, \beta \right) &=
        \frac{\beta^{\alpha}}{\Gamma(\alpha)} x^{\alpha - 1} \textrm{exp}\left(-\beta x\right)  \,.
\end{align*}


%
%

%


\bibliography{main}

\end{document}